\documentstyle[preprint,aps,prb]{revtex}
\begin{document}
\draft
\title{Fractional Quantum Hall Effect in
Bilayer Two Dimensional Hole Gas Systems}
\author{A. R. Hamilton, M. Y. Simmons, F. M. Bolton}
\address{Cavendish Laboratory, Madingley Road, Cambridge CB3 OHE,
 U.K.}
\author{N. K. Patel}
\address{Toshiba Cambridge Research Centre, 260 Science Park, Milton
Road, Cambridge CB4 4WE, U.K.} 
\author{I. S. Millard, J. T. Nicholls, D. A. Ritchie,
and M.  Pepper}
\address{Cavendish Laboratory, Madingley Road, 
Cambridge CB3 OHE, U.K.}
\date{\today}
\maketitle
\begin{abstract}
We have studied the fractional and integer quantum Hall effect in high
mobility double layer 2D hole gas systems. The large hole effective
mass inhibits tunneling, allowing us to investigate the regime in
which the interlayer and intralayer interactions are comparable
without significant interlayer tunneling occurring. As the interlayer
separation is reduced we observe the formation of bilayer correlated
quantum Hall states at total filling factor $\nu$=3/2 and $\nu$=1. We
find that the bilayer $\nu$=3/2 state is rapidly destroyed
by small carrier density imbalances between the layers, whereas the
bilayer $\nu$=1 state evolves continuously into the single layer
$\nu$=1 state.
\end{abstract}


In a bilayer system, new integer and fractional quantum Hall (QH)
states can be observed which have no counterpart in single layer
systems. A good example is the $\nu$=1/2 state: in a clean single
layer system, this filling factor corresponds to a compressible
Fermi liquid,\cite{Willett93} whereas in a bilayer system with total
filling factor 1/2 a well defined fractional QH state can be
observed.\cite{Suen92Eisenstein92} Similarly at $\nu$=1 (i.e.
$\nu$=1/2 in each layer) an interlayer correlated QH state can be
formed, \cite{Suen92Eisenstein92,Hyndman96} although in electron
systems it can be difficult to distinguish this state from a single
particle QH state due to interlayer tunneling. \cite{He93} In
addition to these new quantum Hall effect states, there is also the
possibility of bilayer Wigner crystallisation, which is expected
to occur at larger carrier densities and filling factors than in
single layer systems. \cite{wigner} 

In this paper we present studies of the integer and fractional
quantum Hall effect in bilayer 2D hole systems. We use 2D hole gases
(2DHG) because the large heavy hole mass (0.38$m_e$) allows the
layers to be brought into close proximity without significant
inter-well tunneling occurring. This helps us to distinguish between
QH states arising from the symmetric-antisymmetric gap
($\Delta_{\text{SAS}}$) and those arising from interlayer
interactions. We find that as the layers are brought closer
together, quantum Hall states form at $\nu$=3/2 and $\nu$=1. The
former is the electron-hole conjugate of the $\Psi_{331}$ $\nu$=1/2
bilayer state.\cite{Suen94} Temperature dependence measurements
confirm that the latter is also a correlated state, $\Psi_{111}$,
rather than being due to the single particle tunneling gap
$\Delta_{\text{SAS}}$. However, despite their common origins, we
find that these two states have a very different response to carrier
density imbalance between the two wells. 
 
The samples were grown by MBE on $(311)A$ oriented GaAs
substrates, and consist of two modulation
doped 150{\AA} quantum wells, separated by
Al$_{0.33}$Ga$_{0.67}$As barriers of thickness
$d_B$=20{\AA}, 25{\AA} and 35{\AA}. All of the samples have very
similar carrier densities and exceptionally high
mobilities, with
$p_s{=}1{\times}10^{11}\text{cm}^{\text{-2}}$ and
$\mu{=}1{\times}10^6\text{cm}^2\text{V}^{\text{-1}}\text{s}^{\text{-1}}$
in each well. Measurements were performed with the
samples mounted in the mixing chamber of a dilution
refrigerator with a base temperature less than 30mK.
Temperature dependence measurements were performed using
a calibrated resistor mounted in a flux cancelled region
of the mixing chamber. Standard low frequency ac lockin
techniques were used with measurement currents less than
2nA to avoid heating effects. 

A Schottky front-gate was used to tune the carrier density and
charge asymmetry in the quantum wells. Fig.~\ref{fig1} shows the
carrier densities ($p_s$) in the two layers for the $d_B$=25{\AA}
sample, obtained from Fourier transforms of the low field
($B{<}0.9$T) magnetoresistance oscillations. The sum of these
densities agrees well with the carrier density obtained from the low
field Hall resistance. For gate voltages $V_g{>}0.3$V only the lower
hole gas is occupied. As $V_g$ is reduced the upper 2DHG starts to
occupy, partially screening the lower layer from further reductions
in $V_g$. Whilst the carrier density in the
upper 2DHG rises as $V_g$ is reduced, the carrier density in the
lower layer decreases from $1.3{\times}10^{11}\text{cm}^{\text{-2}}$
at $V_g$=0.3V to 1.1$\times10^{11}\text{cm}^{\text{-2}}$ at
$V_g$=-0.2V. This decrease is due to intralayer interactions in the
upper hole gas, which cause it to have a negative
compressibility.\cite{Ying95,Patel96a} This negative compressibility
effect is significantly larger in hole gases than their electron gas
counterparts, due to the large hole effective mass.\cite{Millard96}
When $V_g$=0.05V the carrier density in the two wells is equal, and
the system is on resonance. At this point the two wavefunctions are
no longer localised in the two wells, but become delocalised across
both wells. Despite the small barrier thicknesses in these samples we
were unable to resolve an anticrossing in the Fourier transform data,
which implies that $\Delta_{\text{SAS}}<85{\mu}$eV. This is in
agreement with self-consistent Hartree calculations which find that
$\Delta_{\text{SAS}}$ is approximately 30$\mu$eV and
10$\mu$eV for the $d_B$=25{\AA} and 35{\AA} samples respectively.
These $\Delta_{\text{SAS}}$ values are almost two orders of magnitude
smaller than those obtained for identical electron gas samples,
highlighting the reduction in tunneling brought about by the large
effective mass. 

Magnetoresistance traces at the resonance point are shown for the
$d_B$=35{\AA} and 25{\AA} samples in figure~\ref{fig2}. Great care was
taken to locate this resonance point, as the magnetoresistance data
was extremely sensitive to any carrier density imbalance between the
two wells.  For the $d_B$=35{\AA} sample there is essentially no
tunneling between the two wells, and so the system behaves as two
independent 2D hole gases with equal carrier densities. The
magnetoresistance data therefore looks like that for a single 2D
system with a double degeneracy and only even numerator QH states,
such as $\nu$=4/3, 2, 10/3, 4\dots are observed, corresponding to
$\nu$=2/3, 1, 5/3, 2\dots in each well. The only exception to this is
a weak feature at $\nu$=3/2 ($B$=6T), which will be discussed
later. Reducing $d_B$ to 25{\AA} (fig.~\ref{fig2}b) increases
tunneling between the two wells, allowing the energy gap
$\Delta_{\text{SAS}}$ between the symmetric and antisymmetric
eigenstates to be resolved. At very low magnetic fields ($B{<}0.2$T) a
weak beating is seen in the magnetoresistance traces due to the
slightly different carrier densities in the symmetric and
antisymmetric subbands. At slightly higher magnetic fields odd index
QH states, such as $\nu$=5, 7, 9\dots become resolved, due to the
$\Delta_{\text{SAS}}$ gap (inset to fig.~\ref{fig2}b). The behaviour
of the odd index QH states at larger magnetic fields is more complex;
the $\nu$=3 state is extremely weak, but the $\nu$=1 state is stronger
again. The feature at $\nu$=3/2 has also become much stronger.

To determine whether the $\nu$=odd QH states in the
$d_B$=25{\AA} sample are due to the $\Delta_{\text{SAS}}$ gap or to
many-body correlations, we have performed temperature dependence
measurements at the resonance point, shown in figure~\ref{fig3}. The
energy gap associated with a given QH state was obtained from fitting
the central linear regions of the Arrhenius plot (inset to
fig.~\ref{fig3}) to the Fermi-Dirac formula
$R_{xx}\propto[1+\exp{(\Delta/2kT)}]^{\text{-1}}$. For low magnetic
fields $\Delta$ increases linearly with $B$, due to Landau-level
mixing,\cite{MacDonald90} and then collapses abruptly at $B$=3T due to
the increasing importance of interlayer
correlations.\cite{MacDonald90,Boebinger92} As the magnetic field is
increased the  number of holes in the highest occupied Landau level
rises, and the average distance between adjacent holes {\it within}
this Landau level, $\langle r_{ij}\rangle$, decreases. As $\langle
r_{ij}\rangle$ becomes smaller, inter-particle Coulomb interactions
become increasingly important. This inter-particle Coulomb energy can
be reduced if the holes change from being delocalised across both
wells to being localised in one well, thereby slightly increasing
$\langle r_{ij}\rangle$. If this reduction in Coulomb energy is larger
than $\Delta_{\text{SAS}}$, tunneling is suppressed, the ground-state
of the system becomes a gapless bilayer state, and all $\nu$=odd 
states disappear. The magnetic field at which this occurs in our
samples is in excellent agreement with the theory of ref.
\onlinecite{MacDonald90}. This magnetic field induced collapse of
the $\Delta_{\text{SAS}}$ gap explains why we do not observe odd
numerator fractions such as $\nu$=5/3 in fig.~\ref{fig2}b, as the
system is doubly degenerate at large $B$ (e.g. $\nu$=4/3 corresponds to
$\nu$=2/3 in each layer\cite{Tdep4/3}). Hence the odd denominator QH
states that we observe above $B$=3T are not due to tunneling, but
must be bilayer correlated states. The state at $\nu$=3/2 is therefore
the electron-hole conjugate of the $\Psi_{331}$ bilayer $\nu$=1/2
state (in bilayer systems electron-hole symmetry transforms
$\nu\rightarrow2{-}\nu$), and the state at $\nu$=1 is the $\Psi_{111}$
bilayer state.\cite{He93} 

We now examine the effects of carrier density imbalance on the
various QH states. Magnetoresistance traces for the $d_B$=25{\AA}
sample, taken at many different gate voltages, are shown as a
greyscale plot in figure~\ref{fig4}. Each magnetoresistance
measurement for a given $V_g$ is represented by a horizontal slice
of the figure. In the greyscale the darkness of the shading is
proportional to the longitudinal resistance $R_{xx}$. Thus the white
regions indicate QH states ($R_{xx}{\rightarrow}0$), where the Fermi
energy lies in the localised states between Landau levels, and the
dark regions correspond the the Fermi energy lying in the extended
states of a Landau level. The numeric labels on the greyscale
identify the total filling factor in the system. Although the
detailed shape of the greyscale is rather complex (and is explained
in more detail elsewhere\cite{Hamilton}), a simple explanation is
given below.

The dashed lines in fig.~\ref{fig4} show the filling factor in the
upper and lower layers,  $\nu_u$ and $\nu_l$, calculated from the
data of fig.~\ref{fig1} using the relation $\nu eB/h=p_s$. For
$V_g{>}0.33$V only the lower layer is occupied, and a normal Landau
fan diagram is seen originating from $V_g$=1.1V. In this regime the
dashed lines therefore mark the positions of the $\nu$=2 and 
$\nu$=3 QH states, which move to larger magnetic fields as the
carrier density is increased by reducing $V_g$. We can also see the
evolution of the fractional QH effect in the lower layer, with
states at $\nu$=1/3, 2/3, and 3/5 becoming more clearly resolved
as $p_s$ increases.

At $V_g$=0.33V the upper layer starts to occupy, and a second Landau
fan emerges. As shown in fig.~\ref{fig1}, once the upper layer
is occupied it partially screens the lower layer from changes in
$V_g$, so the Landau levels associated with the lower layer remain
at almost constant magnetic fields.  For example if we follow
$\nu_l$=3 we see QH states of the bilayer system (indicated by the
white regions where $R_{xx}{\rightarrow}0$) at total filling factor
$\nu=\nu_u{+}\nu_l=4$, 5, 6\dots which correspond to $\nu_l$=3 and
$\nu_u$=1, 2, 3\dots respectively. Away from resonance therefore QH
states form whenever the Fermi energy lies in the localised states
in both layers, as indicated by the intersections of the dashed
lines.

When the two wells are on resonance (at $V_g$=0.05V, indicated by
the arrow) the wavefunctions are delocalised across the two wells,
and we can no longer identify the individual filling factors in the
wells. At this point we not only see $\nu$=even QH states at the
intersections of the dashed lines, as we would for two uncoupled 2D
systems (c.f. fig.~\ref{fig2}a), but also states at $\nu$=odd, due
to the SAS gap (c.f. fig.~\ref{fig2}b).

This simple assignment of filling factors to the two layers can also
be used to explain some features of the higher $B$ data: the
$\nu$=4/3 and $\nu$=5/3 states at $B$=5T correspond to ($\nu_l$=1,
$\nu_u$=1/3) and ($\nu_l$=1, $\nu_u$=2/3) respectively. The Coulomb
interaction driven collapse of $\Delta_{\text{SAS}}$ is responsible
for the absence of $\nu$=3 and $\nu$=5/3 at resonance, although we
see that if the system is taken off resonance $\nu$=4/3
($\nu_l$=$\nu_u$=2/3) is destroyed and $\nu$=5/3 re-emerges.

The most striking feature of the high field data however is the
continuous evolution of the $\nu$=2 and $\nu$=1 from single layer
states in the lower layer ($V_g{>}$0.33V) to bilayer states at
resonance. We believe these effects, which have also been seen in
bilayer electron systems,\cite{anomalousQH} are caused by
charge redistribution between the two wells with
applied perpendicular magnetic field. Although $B$ has little effect
on the wavefunctions, the discrete density of states in the system
means that in order to minimise the Fermi energy it may be
energetically favourable to preferentially populate one energy
level.\cite{Kelly91} This continual redistribution of charge will be
enhanced at high $B$, where the Landau level spacing is large. Thus
we see the single layer $\nu_l$=2 evolves into the bilayer $\nu$=2,
with $\nu_l$=$\nu_u$=1, and the single layer $\nu_l$=1 evolves into
the bilayer correlated $\Psi_{111}$ $\nu$=1.\cite{Lay94} However we
are unable to explain why the $\nu$=1 state rapidly disappears as we
move beyond resonance (i.e. $V_g{<}$-0.01V). 

In comparison with the behaviour of $\nu$=1, figure~\ref{fig5} shows
the behaviour of the $\nu$=3/2 state to carrier density imbalance
(these measurements were performed on a the $d_B$=20{\AA} sample as it
showed this state more clearly). Here we see that at resonance (bold
trace) a clear QH plateau is seen at $\nu$=3/2, with no feature at
$\nu$=5/3. Even moving slightly off resonance causes the $\nu$=3/2
state to weaken, with the associated $\rho_{xx}$ minima disappearing
(not shown), and the small carrier density imbalance ($\approx$ 4\%)
allows the formation of a state at $\nu$=5/3. This is to be expected,
as $\nu$=3/2 is a bilayer correlated state, whereas the $\nu$=5/3
state seen here corresponds to $\nu$=1 in one layer and $\nu$=2/3 in
the other.
 
In summary we have presented the first studies of the fractional QH
effect in bilayer hole systems. We find that as the layer separation
is reduced QH states form at $\nu$=1 and $\nu$=3/2. Temperature
dependence measurements have demonstrated that these states are
correlated bilayer states. We find that the $\nu$=3/2 state is rapidly
destroyed by a carrier density imbalance between the layers,
consistent with its bilayer origin. The behaviour of the $\nu$=1 state
with carrier density imbalance is complicated by the existence of a
single layer $\nu$=1 state: the bilayer correlated $\nu$=1 state
evolves continuously into the single layer $\nu$=1 state with
increasing carrier density imbalance.  Although a simple model of two
non-interacting layers is sufficient to describe the evolution of the
QH effect with carrier density imbalance at low magnetic fields it
cannot describe the more complex behaviour we observe at higher fields
(particularly for $\nu$=1 and $\nu$=2), which warrants further
investigation.

This work was funded by EPSRC (U.K.).


\begin{figure}
\caption{Carrier density as a function of gate voltage for the
$d_B$=25{\AA} sample. At $V_g$=0.05V the carrier density in the two
wells is equal, and at this symmetric point the system is on
resonance. The solid lines are a guide to the eye.} 
\label{fig1}
\end{figure}

\begin{figure}
\caption{Hall ($R_{xy}$) and magnetoresistance ($R_{xx}$) traces for the
$d_B$=35{\AA} and $d_B$=25{\AA} samples at resonance. The insets show
the low field ($B{<}1.5$T) magnetoresistance. $R_{xx}$ traces are at
$T{<}30$mK; $R_{xy}$ traces are at (a) $T$=100mK and (b) $T$=80mK.}
\label{fig2}
\end{figure}

\begin{figure}
\caption{Activation energies of the $\nu$=odd QH states for the
$d_B$=25{\AA} sample at resonance, as a function of magnetic field.
The solid line is a guide to the eye. The inset shows typical
Arrhenius plots, for $\nu$=7 and $\nu$=1, together with Fermi-Dirac
fits from which the activation energies were obtained.} 
\label{fig3}
\end{figure}

\begin{figure}
\caption{Greyscale plot of the longitudinal resistance $R_{xx}$ as a
function of magnetic field $B$ and gate voltage $V_g$ for the
$d_B$=25{\AA} sample at $T{<}30$mK. The depth of
shading is proportional to $R_{xx}$, with white regions denoting QH
states where $R_{xx}{\rightarrow}0$. The resonance point is
indicated by an arrow. The dashed lines show the filling factor in
the upper and lower layers, $\nu_u$ and $\nu_l$, calculated from
the data of fig. 1.} 
\label{fig4}
\end{figure}

\begin{figure}
\caption{Hall resistance $R_{xy}$ as a function of magnetic field at
different gate voltages near resonance for the $d_B$=20{\AA} sample at
$T{<}30$mK. At resonance (bold trace) a strong QH plateau is observed
at $\nu$=3/2, quantised to within 4\% of 2/3 $e^2/h$. The other traces
have been horizontally offset for clarity, with each offset trace
corresponding to an interlayer charge transfer of $\Delta
p_s=6.4{\times}10^8{\text{cm}}^{-2}$.}
\label{fig5}
\end{figure} 


\begin{references}
\bibitem{Tdep4/3}This is confirmed by temperature dependence
measurements, which show that the bilayer $\nu$=4/3 activation gap
is identical to the  $\nu$=2/3 gap measured at the same $B$ when
only the lower layer is occupied.
\end{references}
\end{document}